# H$_2$D$^+$ IN THE HIGH MASS STAR-FORMING REGION CYGNUS-X

T. Pillai[1,2], P. Caselli[3], J. Kauffmann[4,2], Q. Zhang[2], M. A. Thompson[5], D.C. Lis[1]

## ABSTRACT

H$_2$D$^+$ is a primary ion which dominates the gas-phase chemistry of cold dense gas. Therefore it is hailed as a unique tool in probing the earliest, prestellar phase of star formation. Observationally, its abundance and distribution is however just beginning to be understood in low-mass prestellar and cluster-forming cores. In high mass star forming regions, H$_2$D$^+$ has been detected only in two cores, and its spatial distribution remains unknown. Here we present the first map of the *ortho*-H$_2$D$^+$ $J_{k^+,k^-} = 1_{1,0} \rightarrow 1_{1,1}$ and N$_2$H$^+$ 4-3 transition in the DR21 filament of Cygnus-X with the JCMT, and N$_2$D$^+$ 3–2 and dust continuum with the SMA. We have discovered five very extended ($\leq 34000$ AU diameter) weak structures in H$_2$D$^+$ in the vicinity of, but distinctly offset from embedded protostars. More surprisingly, the H$_2$D$^+$ peak is not associated with either a dust continuum or N$_2$D$^+$ peak. We have therefore uncovered extended massive cold dense gas that was undetected with previous molecular line and dust continuum surveys of the region. This work also shows that our picture of the structure of cores is too simplistic for cluster forming cores and needs to be refined: neither dust continuum with existing capabilities, nor emission in tracers like N$_2$D$^+$ can provide a complete census of the total prestellar gas in such regions. Sensitive H$_2$D$^+$ mapping of the entire DR21 filament is likely to discover more of such cold quiescent gas reservoirs in an otherwise active high mass star-forming region.

*Subject headings:* ISM: abundances — ISM: clouds — ISM: molecules — ISM: structure — radio lines: ISM — stars: formation – ISM: individual (Cygnus-X)

## 1. INTRODUCTION

In their evolution toward the formation of a young stellar object, low–mass cloud cores change their density distribution until the so–called "pivotal" state is reached (e.g. Shu et al. 1987). This stage is considered the starting point for protostellar accretion. Thus, it is crucial to find and study objects close to the "pivotal" stage to unveil the initial conditions in the process of star formation. We believe to have found this class of objects: they are the so–called *prestellar cores*, characterised by central densities of $\sim 10^6$ cm$^{-3}$, centrally concentrated density profiles, and with clear signs of contraction in the central few thousand AU (Caselli et al. 2002a,b; Crapsi et al. 2004, 2005). They stand out in having large degrees of CO freeze–out and deuterium fractionation (Bacmann et al. 2003). In particular, they show strong emission in the $J_{k^+,k^-} = 1_{1,0} \rightarrow 1_{1,1}$ line of *ortho*-H$_2$D$^+$ in the case of L1544, the prototypical low-mass prestellar core (Caselli et al. 2003). In this object, the H$_2$D$^+$ line has been mapped by Vastel et al. (2006) and found to be extended over a region of $\sim$5000 AU in projected radius. A H$_2$D$^+$ survey was conducted in low–mass cores (Caselli et al. 2008) and revealed a decrease in fractional abundance of *ortho*-H$_2$D$^+$ in protostellar cores as compared to starless cores.

To reproduce the past observations towards low-mass cores, which imply abundances of H$_2$D$^+$ close to $10^{-9}$ w.r.t. H$_2$, one needs to have species such as CO (which destroy H$_2$D$^+$) almost completely frozen onto dust grains (e.g. Walmsley et al. 2004; Roberts et al. 2004). This can be achieved only if the dust temperature is low ($T_{\rm dust} <$ 20 K) and the density is large ($n[$H$_2] \geq 10^5$ cm$^{-3}$). At these conditions, all species quickly condense out (with time scales $\sim 10^9$ yr $(n[$H$_2]/$cm$^{-3})^{-1}$, leaving mainly He, H$_2$ and its daughter species (in particular H$_3^+$ and its deuterated forms, including H$_2$D$^+$) in the gas phase. Therefore, H$_2$D$^+$ observations are a unique tool to trace and study the chemistry and kinematics of the high density nucleus of prestellar cores, the future stellar cradle.

That being said about low-mass star formation, the early phase chemistry in massive star forming regions is still poorly understood, mainly because the study of these regions have so far focused on molecular material associated with high mass protostellar objects (e.g. Beuther et al. 2002), which are actively altering gas and dust properties. Based on our wide-field ($10 \times 10$ arcmin) census of clouds in 32 massive star-forming regions, harboring UCH II regions, we discovered a new sample of massive pre-protocluster cores. Our program is known as SCAMPS (the SCUBA Massive Pre/Protocluster core Survey; Thompson et al. 2005). Many of the sources are seen as dark patches in infrared images of the region. As a result these clumps must have dust temperatures below 30 K (as evidenced by mid-IR upper limits) and have masses of a few 100 to 1000 M$_\odot$. These mm-peak/mid-infrared absorption sources are very promising candidates for being high mass pre-protocluster cores.

Electronic address: tpillai@astro.caltech.edu
[1] California Institute of Technology, Cahill Center for Astronomy and Astrophysics, Pasadena, CA 91125, USA: Current Address
[2] Center for Astrophysics, 60 Garden Street, Cambridge, MA 02138, USA where a significant part of this work was done
[3] School of Physics and Astronomy, University of Leeds, Leeds LS2 9JT, UK
[4] Jet Propulsion Laboratory, California Institute of Technology, 4800 Oak Grove Drive, Pasadena, CA 91109, USA
[5] Centre for Astrophysics Research, Science and Technology Research Institute, University of Hertfordshire, Herts AL10 9AB, UK

We started investigating these pre-protocluster cores in more detail by observing them in $NH_3$ with the Effelsberg 100 m telescope, as well as in deuterated $NH_3$ with the IRAM 30m telescope(Pillai et al. 2007). On average, the derived temperatures are around 15 K compared with average temperatures toward the previously studied more evolved High Mass Protostellar Objects (HMPOs, Sridharan et al. 2002) of 22 K. Finally, depletion and very large deuterium fractionation, $[NH_2D]/[NH_3]$ $\sim 0.2$, are found (Pillai et al. 2007, Pillai et al. 2011), an enhancement of $\approx 10^4$ over the cosmic D/H ratio. With gas masses of the order of a few hundred $M_\odot$, high deuteration, depletion, and cold temperatures, these sources feature many characteristics expected for massive pre-protocluster cores. Similar large values of the deuterium fraction have been found toward quiescent cores embedded in Infrared Dark Clouds (IRDCs; Chen et al. 2010, Fontani et al. 2011). However see also other recent deuteration studies in IRDCs (Miettinen et al. 2011, Sakai et al. 2012).

With the JCMT and SMA we observed the ortho–$H_2D^+(1_{1,0}-1_{1,1})$, $N_2D^+$ 3–2, and $N_2H^+$ 4–3 line towards three cores from our sample. Here we report detection of the ortho–$H_2D^+(1_{1,0}-1_{1,1})$ line and its large-scale emission in one of the three cores (the only core where ortho–$H_2D^+$ was detected) and subsequent mapping of the deuterated region in $N_2D^+$ 3–2 line with SMA.

## 2. SOURCE SELECTION

The JCMT observations were initially targeted towards three objects from the SCAMPS survey with the brightest $NH_2D$ detections. In total 23.8 hours were spent on the project on the three sources. After few hours of integration on each of the three targets which did not result in a detection, we decided to perform deeper integration on G81.74+0.59, since we detected the brightest $NH_2D$ line towards it in our previous IRAM 30m survey (Pillai et al. 2007). Here, we only discuss the observations towards G81.74+0.59. The maps presented here correspond to clumps along the DR21 filament $\sim 2'$ north of the well known DR21(OH) massive young stellar object. The DR21 filament itself is the most massive region in the entire complex. Our mapped region coincides with three bright 1.3 mm MAMBO dust continuum peaks as revealed in a large scale survey of Cygnus-X (Motte et al. 2007). For reference, this region encompasses the massive dense core N53 of Motte et al. survey (see also Bontemps et al. 2010, Csengeri et al. 2011) and FIR 1/2/3 in infrared studies of this region (Chandler et al. 1993, Kumar et al. 2007). The embedded population identified by Kumar et al. is shown as stars in Fig.1.

## 3. SMA AND JCMT OBSERVATIONS

JCMT $N_2H^+$ and $H_2D^+$ observations were carried out in service mode in several sessions starting August 2007 to August 2008 with the HARP instrument on JCMT. HARP is the heterodyne multi-pixel array at JCMT equipped with 16 receptors separated by 30 arcsec with a foot print of 2 arcmin. The beam size of each receptor is $\approx 14''$. Nyquist sampled maps were achieved using a jiggle pattern with a position-switch to move to the reference. During most of our observations, 14 of the 16 receptors were working. After finding evidence of emission at the edge of our map from the initial observing session, we shifted the map center for the subsequent sessions. The pointing accuracy was on average $< 3''$, and the median system temperature at the $H_2D^+$ frequency was 579 K. The `smurf` software provided as a part of the Starlink package was used to regrid the data, mask the bad receptors, construct 3D data cubes, and co-add the scans. The data were then converted to fits and analysed in CLASS [6].

The Submillimeter Array (SMA[7]) $N_2D^+$ and continuum observations were made with all eight antennas in the most compact configuration (sub-compact) in two tracks in August and September 2008. The observations were done in track sharing mode with several sources per track. For G81.74+0.59, we had four pointings separated at less than half a (primary) beam spacing to cover the same region as the JCMT HARP footprint at 230 and 279 GHz respectively. The 230 GHz receiver was tuned to the CO (2–1) line in the spectral band s13 of the USB for the $N_2D^+$ observations. We used a non-uniform correlator configuration to have a higher spectral resolution of $\sim 0.2 \text{ km s}^{-1}$ for the chunks covering the $N_2D^+$ and CO line. The rest of the correlator was set to $\sim 0.4 \text{ km s}^{-1}$. The data were taken under very good weather conditions ($< 1$mm water vapor), and the typical system temperatures were 100 K. The gain calibrator was MWC349a, $\sim 2\,°$away from the target position. Uranus and 3C454.3 were used as the flux and bandpass calibrator, respectively. Flux calibration is based on SMA flux monitoring of the observed planet and is estimated to be accurate to within 20%. The primary beam at 230 GHz is $\sim 51''$. The data calibration was done in MIR, an IDL based calibration package, and then exported to MIRIAD and GILDAS for further data reduction which included continuum subtraction, and imaging. The center of the final image of the mosaic, the 1 $\sigma$ rms noise level achieved in the continuum images, and the resulting synthesized beam for all observations is listed in Table 1.

## 4. RESULTS

### 4.1. $H_2D^+$ Clumps

The detection of JCMT $H_2D^+$is at a modest signal-to-noise (S/N) with five confirmed S/N peaks (5, 6, 7, 7 and 9 $\sigma$) for a smoothed velocity resolution of $\sim 0.2 \text{ km s}^{-1}$, therefore we obtained only source candidates. In Fig.1, the S/N map obtained (after baseline subtraction) is shown. To increase the S/N for the peak positions, we smoothed the original JCMT map using a symmetrical Gaussian PSF with a full-width at half-maximum (FWHM) of two pixels, evaluated over a length of 10 pixels. 80% of the pixels in the inner $90 \times 90$ ''of the map shown in Fig.1 have rms noise between 0.013 and 0.021 K. This resulted in our final map presented here with a FWHM beam of 19.7 ''. The $> 3\sigma$ detections in the S/N map were then confirmed by visualisation of spectra in the smoothed map, and subsequent S/N analysis on Gaussian fits to the spectra. Similarly, three other $> 3\sigma$ detections (also seen in Fig.1 for example the contour between P4, P1 and P3) were rejected by visualisation of spectra. These might well be true detections, however requires validation with deeper observations. Following these steps, we have confirmed five reliable detections (labelled P1 to P5) whose positional offsets and Gaussian fit parameters (obtained with CLASS) is tabulated in Table 2. The N$_2$H$^+$ (4-3) line parameters for data smoothed to the same spatial resolution is also given in the same table. The spectra towards the five positions for both lines are shown in Fig.2. The spectra have been converted from antenna temperature to main brightness temperature for a main beam efficiency of 0.63 (from the JCMT HARP web page).

---

[6] http://www.iram.fr/IRAMFR/GILDAS/
[7] The Submillimeter Array is a joint project between the Smithsonian Astrophysical Observatory and the Academia Sinica Institute of Astronomy and Astrophysics, and is funded by the Smithsonian Institution and the Academia Sinica.

### 4.2. Spectral Line Features

The H$_2$D$^+$, N$_2$H$^+$ and N$_2$D$^+$ spectral line parameters towards the five H$_2$D$^+$ peaks are reported in Table 2. We also provide the non-thermal line width for these species after taking away the contribution due to thermal broadening at 15 K.

The H$_2$D$^+$ line towards even the brightest core (say P1) at offset (37 ″, 44 ″) is only modest at 0.18 K. Our brightest line is still fainter than the weakest H$_2$D$^+$ detection towards low-mass starless and prestellar cores reported so far (for a complete list see Caselli et al. 2008). The LSR velocity towards the northern sources (P1, P2, P3) is systematically offset from those of the southern cores (P4, P5). We find a range in line widths of $0.8 < \Delta v/\mathrm{kms}^{-1} < 1.6$. The thermal line width (FWHM) at 15 K is 0.4 km s$^{-1}$. Clearly, the line width has a significant non-thermal component as expected in such protocluster environments (also given in Table 2). Though it is difficult to assess the line profiles at the given S/N, a double-peaked profile is observed towards P4. The line asymmetry might be due to different velocity components along the line of sight, absorption in the core envelope or infall motions (as argued by van der Tak et al. 2005 where a similar H$_2$D$^+$ profile is observed in L1544). Clearly, high spatial resolution observations are required to clarify this.

The 4–3 line of N$_2$H$^+$ line is very bright (> 2 K) and broad ($\geq$ 1.5 km s$^{-1}$) towards all the five offset positions. This transition has over 40 hyperfine lines within < 5 km s$^{-1}$. The N$_2$D$^+$ 3–2 transition also has hyperfine structure (hfs) spread over 7 km s$^{-1}$. However this is below our detection limit for all H$_2$D$^+$ peaks. Also, the S/N for N$_2$D$^+$ spectra towards H$_2$D$^+$ peaks is poor. Therefore, we refrain from any interpretation of the N$_2$D$^+$ spectra. H$_2$D$^+$ is much lighter than N$_2$D$^+$ and N$_2$H$^+$ (factor of 7) and hence the thermal line broadening is significant. The thermal line width (FWHM) at 15 K is 0.41 km s$^{-1}$ for H$_2$D$^+$, much larger than 0.15 km s$^{-1}$ for N$_2$H$^+$ and N$_2$D$^+$. However, in Cygnus-X, the non-thermal contribution is still dominant. As listed in Table 2, the intrinsic (i.e. taking the hfs structure into account) line widths of three species are different. Excluding P4 (where the N$_2$H$^+$ and H$_2$D$^+$ line widths are equal), H$_2$D$^+$ line width is lower than that of N$_2$H$^+$ and might imply that deuterated molecules preferably trace quiescent gas. The mean difference in line width for the H$_2$D$^+$ – N$_2$H$^+$pair is 0.52($\pm$0.48). i.e. the higher N$_2$H$^+$ line width with respect to H$_2$D$^+$ is only marginally significant.

The N$_2$H$^+$ LSR velocity is consistently red shifted relative to that of H$_2$D$^+$ and magnitude of this shift varies from 0.2 to 0.6 km s$^{-1}$.

### 4.3. H$_2$D$^+$, N$_2$D$^+$, N$_2$H$^+$ and Thermal Dust Continuum

The dust continuum emission at high (1.3 mm SMA, $\sim$ 3.8 ″) and coarser (1.3 mm MAMBO, $\sim$ 11 ″) angular resolution is shown in panels A and B of Fig.1. The MAMBO clumps fragment further with significant structure all along the filament. The subsequent panels in the figure show the gas distribution in three different tracers: H$_2$D$^+$, N$_2$D$^+$ and N$_2$H$^+$. There is very good correlation between thermal dust continuum and N$_2$H$^+$ emission with the brightest N$_2$H$^+$ emission coinciding with the bright MAMBO peaks (panel D). However, there is no such one to one correlation of dust continuum with either N$_2$D$^+$ (panel C) or H$_2$D$^+$ emission (panels B, C and D).

H$_2$D$^+$ is concentrated in five clumps distributed around the massive protocluster (panel B Fig.1). However, no H$_2$D$^+$ emission is seen towards any of the bright and dense dust cores seen with MAMBO/SMA. Conversely, no compact continuum is detected for H$_2$D$^+$ cores. The H$_2$D$^+$ distribution coincides with our SMA N$_2$D$^+$ 3–2 observations, however their peaks do not. Instead the N$_2$D$^+$ cores are distributed in further clumpy structures offset and partially surrounding the H$_2$D$^+$ peaks. Furthermore, there is one H$_2$D$^+$ peak (P1) with no associated N$_2$D$^+$ emission, and several N$_2$D$^+$ cores with no corresponding H$_2$D$^+$ emission. As seen in the fourth panel of Fig.1, there is always extended N$_2$H$^+$ emission associated with H$_2$D$^+$ cores.

### 4.4. ortho-H$_2$D$^+$ and N$_2$H$^+$ abundance

The ortho-H$_2$D$^+$ column density for the different cores have been estimated following Vastel et al. 2006 from the main beam brightness temperature, line width and assuming an excitation temperature ($T_{\mathrm{EX}}$) of 10 K. The H$_2$D$^+$ line is optically thin in all cores ($\ll$ 1) and the assumed $T_{\mathrm{EX}}$ also enters the optical depth estimation following equation 4 of Vastel et al. 2006. Therefore, we note that the derived H$_2$D$^+$ column density crucially depends on the assumption of excitation temperature. A 30% decrease in $T_{\mathrm{EX}}$ lower than 10 K is poorly justified. Pillai et al. (2007) estimate a ammonia rotational temperature of 18 K in the vicinity of P4 (within 40″). The presence of H$_2$D$^+$ emission together with the lack of 24$\mu$m emission towards the H$_2$D$^+$ cores rules out much higher gas temperatures ($\gg$ 20 K). In their H$_2$D$^+$ sample of prestellar and protostellar cores, Caselli et al. find on average 3 K higher kinetic temperatures and higher $T_{\mathrm{EX}}$ for their protostellar cores as their prestellar cores (8 K vs 11 K). Friesen et al. 2010 also find a higher (than prestellar cores) $T_{\mathrm{EX}}$ of 12 K to be appropriate for their low-mass cluster-forming core in Ophiucus.

MAMBO 1.3 mm data smoothed to the resolution of the final JCMT data have been used to determine H$_2$ column density for a 15 K kinetic temperature following Kauffmann et al. (2008). We have used a dust opacity of 0.0102 cm$^2$/g for dust grains with thin ice mantles and gas density n(H) = $10^6$ cm$^{-3}$as in Ossenkopf & Henning (1994). The ortho-H$_2$D$^+$ column density and the resulting H$_2$D$^+$ abundances are reported in Table 3. The ratio between column density towards cores studied here and those observed towards low-mass cores (summarised in Table 3 of Caselli et al. 2008) varies from 1 to 13.

As mentioned before, the $N_2H^+$ hyperfine structure contribute significantly to the broadening of the line. We fit the $N_2H^+$ hyperfines (METHOD HFS in CLASS) and FWHM so obtained is much lower than from a GAUSS fit (Table 2, Gerin et al. 2001). The optical depth, and therefore the excitation temperature which is usually otherwise derived based on optical depth and line brightness is uncertain. Hence the choice of excitation temperature is somewhat arbitrary and is based on observations in literature. For example, Fontani et al. (2008) derive excitation temperature close to 6 K value towards $N_2H^+$ cores in a high mass star forming region. Assuming optically thin condition, we have estimated the column density from the integrated intensity of $N_2H^+$ $4-3$ lines following Caselli et al. 2002c for an excitation temperature of 5 K. The total $N_2H^+$ abundance is tabulated in Table 3. The uncertainties given in brackets are formal errors obtained by a Gaussian error propagation.

## 5. DISCUSSION

While some of the results discussed above are intuitively easy to understand, many are not. In this section we attempt to explain these anomalies and thus characterise the physical conditions.

The bright MAMBO sources are actively forming stars and clusters (Kumar et al. 2007), as also evidenced by their mid-infrared (mid-IR) emission. Higher temperatures, and subsequent release of heavy neutrals (CO) from grain mantles destroy $H_2D^+$ and therefore reduce the deuterium reservoir in such bright cores. For example, from $NH_2D$ observations we find clear evidence of the peak being offsets from the bright cores showing star formation activity (Pillai et al. 2011). The colder secondary cores on the other hand are expected to have abundant deuterated species. Are these secondary cores principally seen in $H_2D^+$ emission representative of the prestellar/cluster gas?

### 5.1. Extent of $H_2D^+$ emission

The most surprising element of our observation is the wide-spread emission of $H_2D^+$ with no direct counterpart in previous dust continuum or line measurements (continuum data from Motte et al. 2007 and line data from Csengeri et al. 2011). The $H_2D^+$ emission is instead distributed in cores around the bright continuum cores over a $\sim 1$ pc region. The individual cores have on average diameter of 0.2 pc roughly estimated by fitting ellipses by eye (34000 AU, P4 and P5 are smaller). This is more than twice the extent of $H_2D^+$ emission observed towards the only two regions previously mapped in $H_2D^+$; the low-mass prestellar core in Taurus (L1544; Vastel et al. 2006) and the low-mass cluster-forming core in Ophiucus (OphB2; Friesen et al. 2010).

In the past, $H_2D^+$ was thought to trace the innermost regions (few 1000 AU) of a dense core where all the heavier molecules deplete on to grain mantles with density exceeding $10^6$ cm$^{-3}$ (di Francesco et al. 2007). However, with newer data on the collisional rate of $H_2D^+$ the critical density of the transition has been corrected to $\sim 10^5$ cm$^{-3}$, an order of magnitude lower than previously assumed (Caselli et al. 2008). Combined with the prevalent cold temperature in such dense cores, this implies that the $H_2D^+$ distribution is expected to be much larger ($\gg 7000$ AU). The extent of $H_2D^+$ distribution in OphB2 or L1544 can then be comprehended. However, it is surprising that in Cyg-X this distribution is twice as large. Mapping of the entire DR21 filament is likely to discover more such cold quiescent gas pockets in an otherwise active high mass star-forming region.

### 5.2. Absence of Dust Continuum peaks towards ortho-$H_2D^+$ peaks

Although there is weaker dust emission towards all $H_2D^+$ cores, none of the $H_2D^+$ peaks have been detected as a peak in dust continuum emission with MAMBO as well as SMA. Depletion is often invoked to explain such lack of correlation between dust and gas in different tracers. However, the molecule in question is $H_2D^+$, one of the lightest deuterated ions that should survive depletion. Therefore, we suggest that dust continuum observation is partially misleading, i.e. not tracing mass distribution. We assert this by showing that $H_2D^+$ emission comes from a region of low temperature and high density.

We draw out the density in three different ways. First, let us define a spherical core around a $H_2D^+$ core, for e.g. P1. It has a mass of 137 $M_\odot$ within a radius of $\sim 0.1$ pc from MAMBO observations for a temperature of 15 K. This implies an average density of $5 \times 10^6$ cm$^{-3}$. Second, we extract the intensity of the filament in dust emission towards an intensity minima in between two bright MAMBO cores, considerably offset from the $H_2D^+$ cores and outside its 3 $\sigma$ boundary. Adopting the distance to the extreme ends of two $H_2D^+$ cores boundary as a measure of the size of a typical core, in this case 60 arcsec, we again derive density $(2 \times 10^5$ cm$^{-3})$. The critical density of the observed $H_2D^+$ transition at 20 K is $\sim 1.1 \times 10^5$ cm$^{-3}$. This shows that the density in a region larger than the $H_2D^+$ core by a factor of $\sim 60/20 = 3$ (20 arcsec corresponds to representative $H_2D^+$ core size) is still high enough to get $H_2D^+$ excited. Finally, $N_2H^+$ 4–3 emission is observed all along the filament including towards $H_2D^+$ cores. The critical density [8] of this $N_2H^+$ transition at $3 \times 10^7$ cm$^{-3}$ is two orders of magnitude higher than that of the $H_2D^+$ transition, requiring very dense gas. Clearly not every region (like the envelope) in the filament can have such high densities or the implied high excitation temperature, and the line is likely sub-thermally excited (Evans 1999). We therefore used the non-LTE radiative transfer code RADEX [9] for a simple calculation of the expected $N_2H^+$ 4–3 line brightness for a column density of $10^{15}$ cm$^{-2}$ and line width (FWHM) of 2 km s$^{-1}$ at 20 K for a wide range of densities (guided by $N_2H^+$ line parameters for the $H_2D^+$ cores given in Table 2 – 3). We are thus able to rule out densities below $10^5$ cm$^{-3}$ for the $H_2D^+$ cores.

Next, we look at the maximum temperature of the $H_2D^+$ emitting region. $H_2D^+$ cores are mid-IR-quiet at 24$\mu$m, while the bright MAMBO cores have bright 24$\mu$m point like emission as observed with the MIPS instrument on Spitzer telescope (Spitzer Legacy Pro-gram ID 40184, PI J. Hora), consistent with the notion that these bright objects are massive protostars (Chandler et al. 1993). A precise luminosity of these protostars is difficult to measure because the sources are saturated. Based on a single temperature Spectral Energy Distribution (SED) fit to their Far-Infrared(FIR) and 1.3 mm data, Chandler et al. have determined the luminosity of one of the protostars in the region (known as FIR1) to be ∼ 1860 L$_\odot$ (this value has been corrected for our adopted distance of 1.7 kpc instead of 3 kpc). Given the poor spatial resolution of the IRAS data, the individual protostars would in any case not be distinguished. We find two sources from the Red MSX Source (RMS) survey of a large sample of massive young stellar objects from the MSX point source catalog in our field (Hoare et al. 2005,Urquhart et al. 2008). The survey database at www.ast.leeds.ac.uk/RMS characterises the luminosity of the two sources, G081.7522+00.5906, G081.7624+00.5916 as 2500 and 1000 L$_\odot$ respectively (Mottram et al. 2011). Assuming that FIR1/2/3 are in similar stages of evolution, we adopt 1000 L$_\odot$ as a conservative estimate of the protostellar luminosity. The mass-weighted dust temperature within an aperture of radius $R$ is approximated using Eq. (8) of Kauffmann et al. (2008). That equation assumes a simplified dust temperature dependence on radius and luminosity (i.e., calibrated by Terebey et al. 1993; $T_{\rm d} \propto L^{0.2}/r^{0.4}$), but further requires temperatures ≥ 10 K. Assuming a density profile $\varrho \propto r^{-2}$, these temperatures are then weighed by mass to obtain the mean temperature (e.g., Belloche et al. 2006). Following Eqs. (5), (8) from Kauffmann et al. (2008), the temperature at a radius of 30000 AU (∼ 17 ″) from the protostar would drop to 15 K. Here, the exact value of the radius is not relevant but within reasonable distance of the protostar the dust should be cold, a requisite for H$_2$D$^+$ emission.

---

[8] Derived based on $N_2H^+$ collisional rate from LAMDA database for a temperature of 20 K (Schöier et al. 2005).
[9] van der Tak et al. (2007)

So far we have shown that high density and low temperature — a necessary condition for H$_2$D$^+$ excitation — is satisfied in an extended region around the protostars. Therefore, the cores seen in H$_2$D$^+$ need not necessarily correspond to a 3D maximum in density. To the extent of the validity of our first order calculations here in terms of effect or protostellar luminosity on heating, it is not required to have a density peak towards the H$_2$D$^+$ peaks. If we were to do more elaborate calculations to derive the density structure, then a density peak would have enhanced the cooling within the peak. However, the H$_2$D$^+$ peaks might well correspond to a temperature minima. We intend to investigate this via NH$_3$ measurements.

Why are the H$_2$D$^+$ cores not detected in SMA 1.3 mm dust continuum? The SMA continuum distribution is offset from the H$_2$D$^+$ peaks. Recall however that the H$_2$D$^+$ peaks need not be density peaks. Therefore the fact that the dust peaks are offset is not inconsistent with the H$_2$D$^+$ observations. Even so, it is worth noting that all H$_2$D$^+$ cores are widely associated with SMA dust continuum peaks, i.e. offset from rather than being anti-correlated with continuum peaks. Based on the 1.3 mm MAMBO data we calculated the total mass of the H$_2$D$^+$ cores. We derive the gas mass at 15 K from the 1.3 mm following Kauffmann et al. (2008), using dust opacities at 1.3 mm (0.0102 cm$^2$/g) for dust grains with thin ice mantles and gas density n(H) = 10$^6$ cm$^{-3}$ as in Ossenkopf & Henning 1994, and gas to dust ratio of 100. The mass estimate is uncertain by a factor of 2 mainly due to the uncertain dust opacity. The typical mass of a H$_2$D$^+$ core is ∼ 130 M$_\odot$. However, we find that ≤ 10% mass resides in the cores detected with SMA. The missing matter might be either dense but smooth structures filtered out by an interferometer like the SMA. This is validated by our N$_2$D$^+$ SMA data. It is contaminated by negative bowls (see third panel of Fig 1). This unwanted feature of synthesis array observations is a signature left by very extended objects which dominate the short spacings in the uv plane. Alternatively, a cluster of low mass cores well below the sensitivity limit of the array would also explain the missing matter.

Given that 90% of the dust emission is missing in the SMA data, the lack of a one-to-one correlation between SMA continuum and H$_2$D$^+$ might not be significant. For example, consider the dust emission associated with clump P1. We know that ∼ 90% of the dust emission is filtered out with SMA. If the structure is smooth, then this ∼ 90% emission could be randomly removed from any position in clump P1. Therefore the remaining emission might also be located at a random position within P1. Moreover, the offset between SMA dust and H$_2$D$^+$ peaks is suggestive of density not being the only criteria to have H$_2$D$^+$ peaks where they are observed. If it would have been the case, then naturally the peak with the highest density would be least likely to filtered out by SMA. Therefore we believe that the H$_2$D$^+$ peaks might be dominated by temperature effects. The dust might be significantly cooler than in the surroundings (and definitely cooler than at the bright dust continuum peaks, where protostars are embedded). A similar reasoning was proposed by Di Francesco et al. (2004) who found that interferometric observations of N$_2$H$^+$ actually betrayed the presence of a dense and cold starless core N6 in Oph A with no distinct peak observed in mm dust continuum observations.

We have derived the virial mass of the H$_2$D$^+$ cores following Bertoldi & McKee 1992. For a mean radius of 0.1 pc and FWHM $\Delta v = 1.2\,{\rm km\,s^{-1}}$, the virial mass of a H$_2$D$^+$ core is 30 M$_\odot$. The average virial mass is 28 M$_\odot$, very small compared to the mass derived from MAMBO observations. Therefore these cores are gravitationally super-critical. The line width is highly supersonic (> 1 km s$^{-1}$), for all but one core (P4). The high line-width is another indication of a clumpy substructure within each H$_2$D$^+$ emitting region.

### 5.3. Absence of N$_2$D$^+$ peaks towards ortho-H$_2$D$^+$ peaks

As noted before, while N$_2$D$^+$ emission as observed at high angular resolution with SMA is widely associated with H$_2$D$^+$, the peaks of these two ions however do not coincide. Lack of N$_2$D$^+$ emission towards P2 is at least partially due to the problem of short spacing discussed in Section 5.2. Briefly, if the N$_2$D$^+$ emission is as extended as the observed H$_2$D$^+$ emission, then a significant fraction of this emission is likely to be filtered out by the SMA. We intend to pursue single dish observations to address this issue. Two cases deserve more attention and are discussed here. First, N$_2$D$^+$ is seen to skirt the H$_2$D$^+$ emission in non-uniform clumpy structures except in the case of P2. The critical density for the 3–2 transitions of this line is an order of magnitude higher than our $H_2D^+$ transition. This implies that higher density is a requisite for its excitation. We have also shown that the $H_2D^+$ peaks at low resolution need not be density peaks. The SMA continuum cores however should point us towards the highest density compact cores in the vicinity of $H_2D^+$ emission. It is exactly here that $N_2D^+$ emission is concentrated.

Next, two of the $N_2D^+$ clumps in the south-west part of the region are devoid of $H_2D^+$ emission. It is difficult to understand why there are $N_2D^+$ cores with no $H_2D^+$ emission. Here, however excitation might play a role. Also, given that the SMA data is prone to artifacts due to the uncertainties inherent in interferometric observations, these structures might be spurious. However, we downloaded published and archived $H^{13}CO^+$ 1–0 PdBI taken at similar resolution to our SMA observations. The region is known as MDC N53 (Csengeri et al. 2011). Clearly, there is $H^{13}CO^+$ emission towards all the $N_2D^+$ features including the two features not detected in $H_2D^+$. The peak $N_2D^+$ and $H^{13}CO^+$ line brightness are however much weaker (more than a factor of 4) toward these $N_2D^+$ cores. If the same scaling of line brightnesses is applied to $H_2D^+$, then it is evident why $H_2D^+$ is not detected (our highest $H_2D^+$ S/N is 8).

### 5.4. *Depletion of N-bearing molecules*

We will now discuss the distribution of $N_2H^+$ and $H_2D^+$. At the low temperature of molecular clouds (T< 20 K), $H_2D^+$ is the primary ion for deuterium transfer to other molecules including $N_2$, which produces $N_2D^+$. Therefore, colder cores exhibit high deuteration as seen in the fractionation of molecules like $N_2H^+$, i.e high $[N_2D^+]/[N_2H^+]$. However, at such low temperatures molecules freeze out onto dust grains. This not only includes CO and its derivatives like $HCO^+$ but also N-bearing molecules like $N_2$ (and consequently $N_2H^+$ and $NH_3$, which are both formed by molecular nitrogen). $H_2D^+$ on the other hand can become abundant under such conditions where all heavy species have disappeared (depleted). To test this, we looked at the abundance distribution of $N_2H^+$ over the region. While the average $N_2H^+$ abundance of $H_2D^+$ cores is $1 \times 10^{-8}$, no significant variation from those position offset from $H_2D^+$ cores is found. The highest abundance is found towards the MAMBO bright peaks which are however higher by only a factor of $\sim 2$. Given that our resolution is no finer than 34000 AU, this result is not surprising. Stronger depletion might well be observed on much smaller spatial scales. While the estimated $N_2H^+$ abundance of $\sim 10^{-8}$ might seem significantly higher than that found in low-mass star forming regions with typical abundance of $\sim 10^{-10}$ (Caselli et al. 2002a, Keto et al. 2004), we caution that the somewhat arbitrary choice of $T_{EX}$ (5 K) influences this calculation. The higher column densities characteristic of high mass star forming regions might explain the high abundance, however a $T_{EX}$ higher by 2 K is sufficient to lower the derived abundance by one order of magnitude.

### 5.5. *Deuterium Chemistry*

How does the observed ortho-$H_2D^+$ abundance compare to existing chemical models? Given the poor spatial resolution of our $H_2D^+$ observations, a tailored chemical model is not warranted. Qualitative comparisons can still be made with models that have focused on the $H_2D^+$ chemistry in dark clouds (Walmsley et al. 2004, Flower et al. 2004). To do this, we need to convert the observed ortho-$H_2D^+$ abundance to the total $H_2D^+$ abundance assuming an ortho-to-para (o/p) ratio for $H_2D^+$. Few observations have constrained this value in dark clouds ($\simeq 0.8$ in L1544 Vastel et al. 2006). Assuming an o/p-$H_2D^+$ ratio as in L1544, we get a total $H_2D^+$ abundance of $\sim 3 \times 10^{-11}$. However, if the gas temperature is even a few degree warmer than in low-mass star forming regions, then the o/p-$H_2D^+$ would drop – as shown in Figure 6 of Flower et al. (2004). This is mainly due to dependence of o/p-$H_2D^+$ on o/p-$H_2$, and an increase in ortho-$H_2$ as shown by chemical models (Pagani et al. 1992). Such a drop in the ortho-$H_2D^+$ column density was already noticed by Caselli et al. (2008) toward warmer starless cores and protostellar cores in their sample with higher deuterium fractionation ($R_D$). We do not have temperature or $R_D$ measurements towards all the $H_2D^+$ cores. However, Pillai et al. (2007) estimate a high rotational temperature of 18 K in the vicinity of P4 (within $40''$) and $R_D = 0.34$ where $R_D$ is measured as the ratio of $NH_2D$ to $NH_3$ column density. This corresponds to a o/p-$H_2D^+$ ratio of about 0.6 (from Fig. 6 of Flower et al. 2004). We also compared our ortho-$H_2D^+$ abundance to the dynamical model of Flower et al. (2006) that incorporates deuterium chemistry in the early stages of protostellar collapse, specifically their Figure 6 where ortho-$H_2D^+$ abundance at 10 K is plotted as a function of the gas density during collapse. For a density of $\sim 5 \times 10^5$ cm$^{-3}$ o/p-$H_2$ close to 0.03 is consistent with our observations. An o/p-$H_2$ of $<< 1$ that we indirectly infer is consistent with those measured in nearby low mass star forming regions. Observations of deuterated molecules with its parent molecule for e.g. $DCO^+/HCO^+$ (Maret & Bergin 2007), $N_2D^+ N_2H^+$ (Pagani et al. 2009), non-LTE radiative transfer modeling of $H_2CO$ absorption (Troscompt et al. 2009), abundance ratio of Nitrogen hydrides (Dislaire et al. 2012) have all delivered o/p-$H_2$ of $<< 0.1$. This implies that the para-$H_2$ dominates in molecular clouds. Similar work in high mass star forming region is yet to be done. Detailed modeling should be made with higher resolution data together with tighter constraints on the density and temperature in Cygnus-X.

### 5.6. $H_2D^+$ *emission in high and low-mass star forming regions*

We present here the first map of $H_2D^+$ distribution in a high mass star forming region. Previously, single spectra of $H_2D^+$ have been reported in at least two such regions, OriB9, a 100 M$_\odot$ cold, quiescent core in OrionB (Harju et al. 2006) and $\sim 4\,\sigma$ detection in a core in IRDC MSX G030.88+00.13 (Swift 2009). The *ortho*-$H_2D^+$ abundance in Cygnus-X is in between the values for OriB9 ($\sim 10^{-10}$) and IRDCs $\sim 10^{-13}$. However, Cygnus-X is also intermediate between these two regions in terms of its distance from us. Therefore, the extreme range in $H_2D^+$ abundance might merely reflect the effect of beam dilution. However, the line width of our $H_2D^+$ cores are similar to the core in G030.88+00.13 (0.9 km s$^{-1}$), and a factor of two higher than in OriB9 ($\sim 0.4$ km s$^{-1}$). We also note that due to a lack of continuum data, Harju et al. derive OriB9 abundance based on the known Ammonia abundance in the region (not to-wards the core), and might be offset from its true value. The H$_2$D$^+$ column density in OrionB9 itself is comparable to Cygnus-X. The difference in line width is also reconciled if we account for the different spatial scales probed. Using the line width-size relation with an exponent of 0.4, if we were to scale the line width at a physical resolution of 0.03 pc i.e. 17 $''$ FWHM APEX beam at 410 pc to that at 0.16 pc (13 $''$ FWHM JCMT beam at 1700 pc), the expected line width is even higher than the observed line width. Therefore the physical conditions in Cygnus-X might be similar to those in the quiescent and massive core in Orion.

As noted before, the average ortho-H$_2$D$^+$ abundance in Cygnus-X is a factor of 1– 20 lower than that in low-mass cores (Caselli et al. 2008). Assuming that the true abundance is similar to what is measured in low-mass star forming regions, this might be due to a low filling factor. In that case H$_2$D$^+$ emission would be widespread but concentrated in small regions, with a filling factor of $\sim 0.1$. H$_2$D$^+$ has been mapped in two low-mass star forming regions, Oph B2, a dense core in the low-mass cluster forming region in Ophiucus (Friesen et al. 2010), and L1544, a prestellar core in Taurus (Vastel et al. 2006). The peak H$_2$D$^+$ abundance in L1544 is an order of magnitude higher than for Cygnus-X. Vastel et al. have used a spherically symmetric chemical model incorporating the physical structure of the cloud (density, temperature), and fit the abundance variation of H$_2$D$^+$ as a function of distance from the core center. While the best fit to their H$_2$D$^+$ finds a peak at $10^{-10}$, $\chi_{o-H_2D^+}$ drops to $\sim 10^{-11}$ at $\sim 10000$ AU from the center, still smaller than our typical $\sim 17000$ AU radii. This re-affirms our earlier conclusion that these observations have uncovered extended cold dense gas that were undetected with typical molecular line and dust continuum surveys. Motte et al. (2007) surveyed the entire complex with MAMBO with the goal of identifying high mass cores at the earliest stage of evolution, but did not find high-mass analogs of prestellar dense cores. In the region of overlap with our H$_2$D$^+$ map, a single object (CygX-N53) has been identified as a massive dense core, which is distinctly offset from the H$_2$D$^+$ cores. However, the three massive dense, colder (and younger) cores discovered by this work (P1, P2, P3) are also undetected in line observations (H$^{13}$CO$^+$ and HCN 1–0) of Csengeri et al. 2011.

Now, L1544 is a prestellar core with just enough mass (8 M$_\odot$, Tafalla et al. 1998) to form a single low-mass star. Oph B2 on the other hand is a low-mass cluster forming core. Many of the H$_2$D$^+$ line properties in Cygnus-X are mirrored in Oph B2. Namely, the H$_2$D$^+$ peak is offset from the dust as well as N$_2$D$^+$ peaks. The line widths are higher than in L1544, although still transonic. The emission is also distributed over a larger region. H$_2$D$^+$ core sizes are not constrained by these observations, however the whole emission extends to $\sim 14000$ AU. Naively, the more extreme H$_2$D$^+$ properties in a high mass region like Cygnus-X seems to be scaled up from low-mass clusters.

## 6. SUMMARY

We present the first map of *ortho*-H$_2$D$^+$ $J_{k^+,k^-} = 1_{1,0} \rightarrow 1_{1,1}$ transition in a high mass star forming region. Distributed over a physical scale of 1 pc in at least five clumpy structures, this is also the largest H$_2$D$^+$ map of any star forming region. The detection is weak and shows an anti-correlation with star-formation, wherein H$_2$D$^+$ clearly avoids the active star forming clusters, however is still in their vicinity and has no IR emission of its own. The H$_2$D$^+$ emission is widely associated with dense dust emission as well as N$_2$D$^+$ as observed with SMA at high angular resolution. However, surprisingly the dust and N$_2$D$^+$ peaks are distinctly offset from the H$_2$D$^+$ peaks. While the derived H$_2$D$^+$ abundance is up to an order of magnitude lower than in low-mass star forming regions, the N$_2$D$^+$ and N$_2$H$^+$ column density and abundance is comparable to or even higher than low-mass star forming regions. The lack of local dust counterparts remains to be better understood. Our best reasoning is that H$_2$D$^+$ cores might have significantly clumpy sub-structure with temperature minima at the observed H$_2$D$^+$ peaks that need not be density peaks. We have thus shown that H$_2$D$^+$ is as good and unique a tool in tracing the cold and dense gas in low-mass star forming regions as in such complex massive star forming regions. We have discovered a large reservoir of cold and dense gas undetected by previous surveys that were targeted at identifying the earliest stages of high-mass star formation. This suggests that submm continuum studies alone which are aimed at detecting massive prestellar cores might fail to identify a large population of such cores in earliest stages. Sensitive H$_2$D$^+$ mapping of the entire DR21 filament is likely to discover more of such cold quiescent gas reservoirs in an otherwise active high mass star-forming region.

We are grateful to Dr.Iain Coulson, and the JCMT staff for their kind support with observations, and data handling. The James Clerk Maxwell Telescope is operated by The Joint Astronomy Centre on behalf of the Science and Technology Facilities Council of the United Kingdom, the Netherlands Organisation for Scientific Research, and the National Research Council of Canada. The JCMT data were obtained under the program ID M07BU14. This research was supported by an appointment of J.K. to the NASA Postdoctoral Program at the Jet Propulsion Laboratory, administered by Oak Ridge Associated Universities through a contract with NASA. It was executed at the Jet Propulsion Laboratory, California Institute of Technology, under contract with the National Aeronautics and Space Administration. T.P. acknowledges support from the Combined Array for Research in Millimeter-wave Astronomy (CARMA), which is supported by the National Science Foundation through grant AST 05-40399. T.P. acknowledges support from the SMA Fellowship Program while working on this project.

*Facilities:* JCMT(HARP), SMA, Spitzer (IRAC).

TABLE 1
Continuum sensitivity.

| Source | Frequency GHz | R.A. (J2000) J2000 | Dec. (J2000) J2000 | 1 $\sigma$ noise mJy | resolution arcsec×arcsec, degrees |
|---|---|---|---|---|---|
| G81.74+0.59 | 230 | 20:39:02.97 | 42:25:05.301 | 4.0 | 4.2×3.5, -67 |

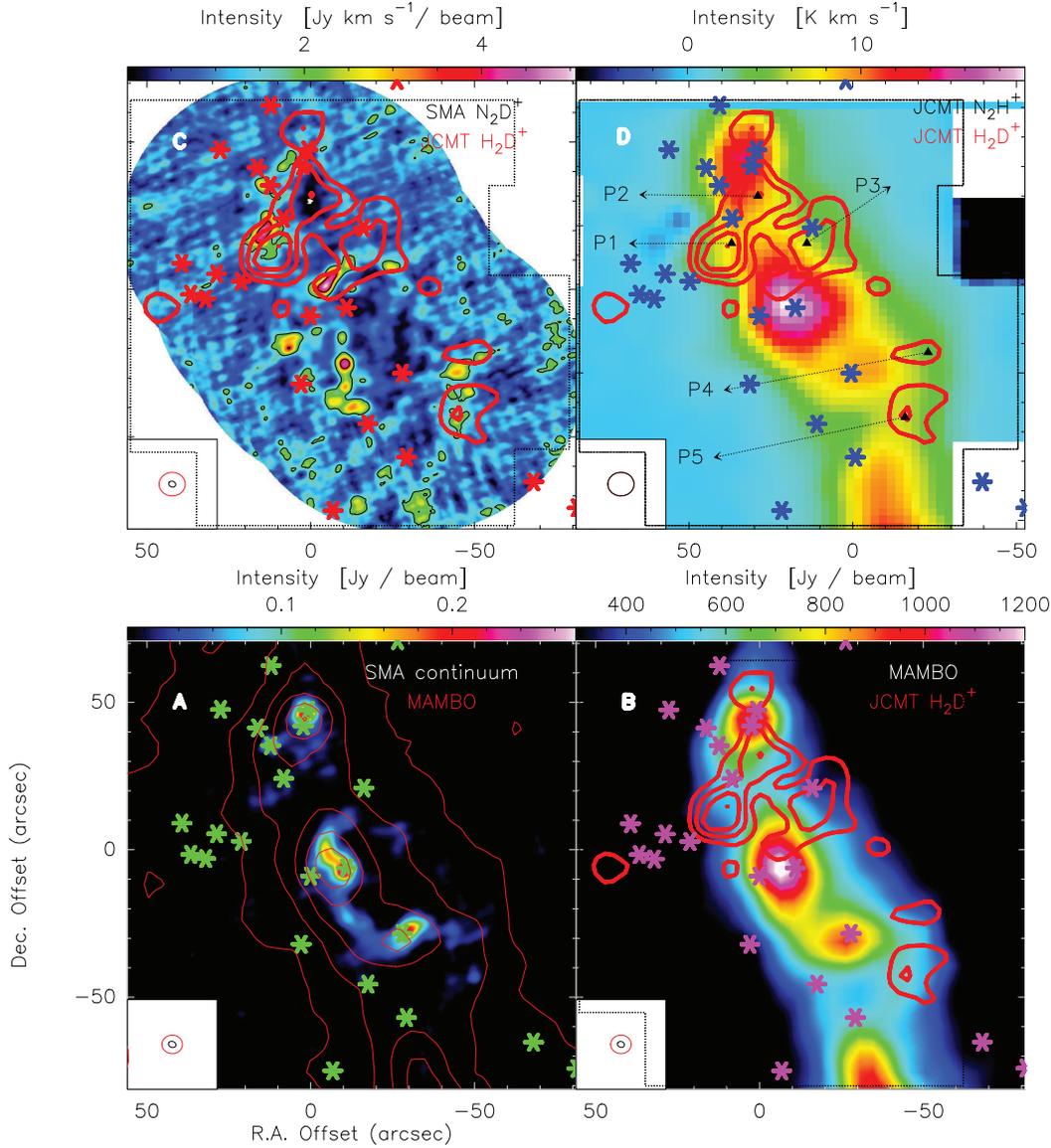

Fig. 1.— *Panel A (from left)*: SMA 1.3mm dust continuum map which is a mosaic of four pointings (color scale), and MAMBO 1.3 mm map (contours, Motte et al. 2007). *Panel B*: MAMBO 1.3 mm dust continuum and H$_2$D$^+$ S/N map as contours (contours, 3$\sigma$ in steps of 1.5$\sigma$ where $0.013 \leq \sigma/K \leq 0.021$). *Panel C*: N$_2$D$^+$ 3–2 moment 0 map (weighted mean intensity integrated from -14 km s$^{-1}$ to -8 km s$^{-1}$.), and JCMT H$_2$D$^+$ S/N map (contours, 3$\sigma$ in steps of 1.5$\sigma$). *Panel D*: N$_2$H$^+$ 4–3 moment 0 map (weighted mean intensity integrated from -10 km s$^{-1}$ to 3 km s$^{-1}$, grey scale), and JCMT H$_2$D$^+$ S/N map (contours, 3$\sigma$ in steps of 1.5$\sigma$). The H$_2$D$^+$ core positions are marked as P'n'. Stars denote the embedded population identified by Kumar et al. (2007).

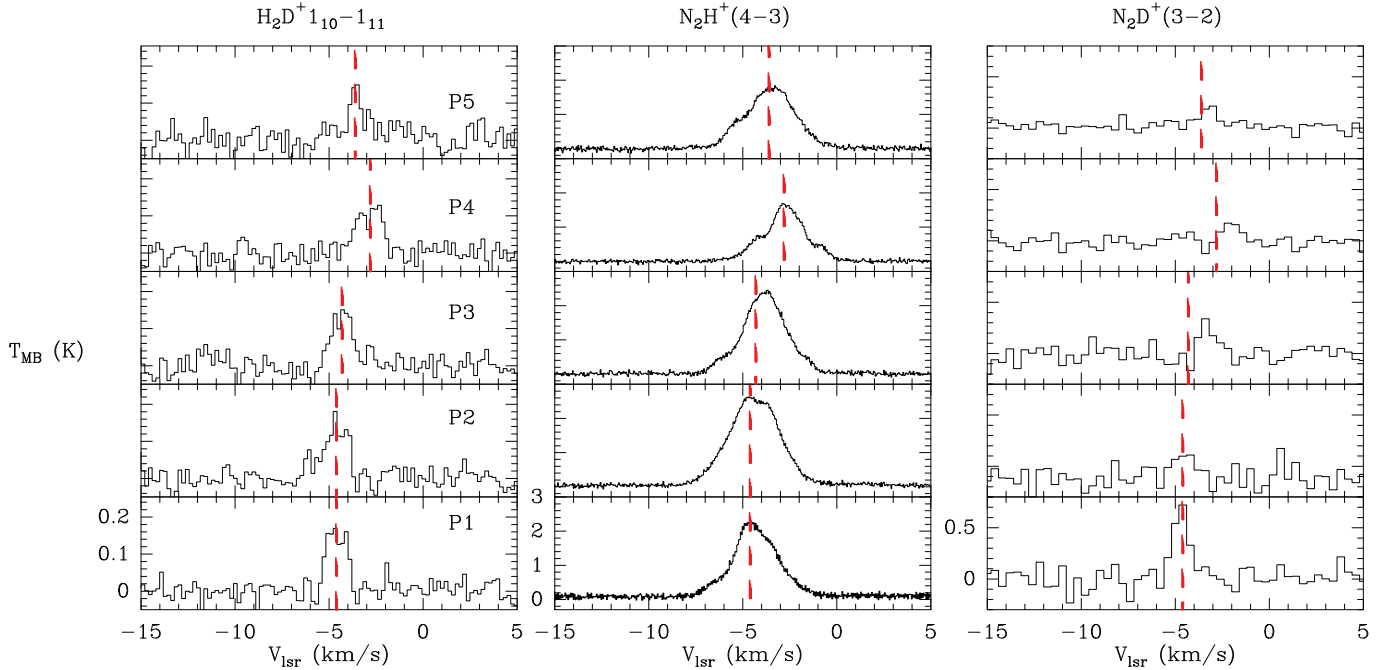

FIG. 2.— *Left Panel*: JCMT spectra of $H_2D^+$ at 372.421 GHz. *Centre Panel*: JCMT $N_2H^+$ 4–3 at 372.672 GHz. *Right Panel*: SMA $N_2D^+$ 3–2 at 231.5 GHz. Spectra for all tracers have been extracted for the same spatial resolution, and the offset positions (in arcsecs) are as indicated in the figure. Note that the SMA spectra have a factor of 2 poorer spectral resolution than the JCMT spectra ($\sim 0.2$ km s$^{-1}$). $H_2D^+$ LSR velocity for each position is indicated as the line of reference in the three panels.

TABLE 2
Gaussian fit parameters from JCMT Observations.

| Source (RA, Dec.) J2000 | Molecule | $T_{MB}$ K | $V_{LSR}$ km s$^{-1}$ | $\Delta v$ km s$^{-1}$ | $\Delta v_{nt}$ km s$^{-1}$ | Velocity Range km s$^{-1}$ | $\int T_{MB}\, dv$ K km s$^{-1}$ |
|---|---|---|---|---|---|---|---|
| P1 | $H_2D^+$ $1_{1,0}$–$1_{1,1}$ | 0.18 (0.02) | -4.6 (0.1) | 1.20 (0.08) | 1.13 | | |
| (20:39:03.7,+42:25:21) | $N_2H^+$ 4–3 | 3.27 (0.03) | -4.2 (0.1) | 1.77 (0.02) | 1.76 | (-8.0,0.0) | 9.65 (0.06) |
| | $N_2D^+$ 3–2 | 0.73 (0.14) | -4.8 (0.1) | 0.82 (0.14) | 0.81 | | |
| P2 | $H_2D^+$ $1_{1,0}$–$1_{1,1}$ | 0.14 (0.02) | -4.6 (0.1) | 1.50 (0.17) | 1.44 | | |
| (20:39:03.1,+42:25:37) | $N_2H^+$ 4–3 | 4.15 (0.02) | -4.2 (0.1) | 2.04 (0.01) | 2.03 | (-8.0,0.0) | 13.72 (0.03) |
| | $N_2D^+$ 3–2 | 0.24 (0.12) | -4.7 (0.2) | 0.85 (0.49) | 0.84 | | |
| P3 | $H_2D^+$ $1_{1,0}$–$1_{1,1}$ | 0.15 (0.02) | -4.3 (0.1) | 1.26 (0.13) | 1.19 | | |
| (20:39:01.7,+42:25:21) | $N_2H^+$ 4–3 | 3.69 (0.02) | -3.7 (0.1) | 1.69 (0.16) | 1.68 | (-8.0,0.0) | 10.79 (0.04) |
| | $N_2D^+$ 3–2 | 0.33 (0.13) | -3.5 (0.1) | 0.67 (0.23) | 0.65 | | |
| P4 | $H_2D^+$ $1_{1,0}$–$1_{1,1}$ | 0.12 (0.02) | -2.8 (0.1) | 1.61 (0.15) | 1.56 | | |
| (20:38:58.3,+42:24:44) | $N_2H^+$ 4–3 | 2.40 (0.02) | -2.6 (0.1) | 1.47 (0.01) | 1.46 | (-8.0,2.0) | 6.93 (0.03) |
| | $N_2D^+$ 3–2 | 0.19 (0.07) | -2.2 (0.1) | 0.69 (0.99) | 0.67 | | |
| P5 | $H_2D^+$ $1_{1,0}$–$1_{1,1}$ | 0.13 (0.07) | -3.6 (0.1) | 0.82 (0.35) | 0.71 | | |
| (20:38:58.9,+42:24:22) | $N_2H^+$ 4–3 | 2.77 (0.02) | -3.4 (0.1) | 1.92 (0.01) | 1.91 | (-8.0,2.0) | 9.30 (0.04) |
| | $N_2D^+$ 3–2 | 0.20 (0.08) | -3.3 (0.1) | 0.74 (0.28) | 0.72 | | |

$\Delta v$ and $\Delta v_{nt}$ are the observed and non-thermal FWHM line width, respectively. $\Delta v_{nt}$ is calculated after accounting for thermal line width at 15 K. Towards position P4, the $H_2D^+$ spectrum shows an asymmetry which if taken as due to two velocity components has line width (FWHM) of 0.88 and 0.68 km s$^{-1}$ at LSR velocity of -3.3 and -2.4 km s$^{-1}$, respectively. The FWHM derived for $H_2D^+$ is from a simple Gaussian fit while that for $N_2D^+$ and $N_2H^+$ is from a hyperfine fit.

TABLE 3
H$_2$D$^+$, N$_2$H$^+$ Column density and abundance.

| Position | N(o−H$_2$D$^+$)[a] (×10$^{12}$) cm$^{-2}$ | $\chi_{\rm o-H_2D^+}$[b] (×10$^{-11}$) | N(N$_2$H$^+$)[c] (×10$^{15}$) cm$^{-2}$ | $\chi_{\rm N_2H^+}$[d] (×10$^{-8}$) |
|---|---|---|---|---|
| P1 | 3.8 (0.5) | 1.9 | 2.0 (0.01) | 1.0 |
| P2 | 3.7 (0.7) | 1.4 | 2.8 (0.01) | 1.1 |
| P3 | 3.3 (0.6) | 1.7 | 2.2 (0.01) | 1.1 |
| P4 | 3.4 (0.7) | 2.1 | 1.4 (0.01) | 0.9 |
| P5 | 1.9 (1.3) | 0.9 | 1.9 (0.01) | 0.9 |

[a]The column density determined for excitation temperature, $T_{\rm EX}$= 10 K.
[b]Fractional *ortho*-H$_2$D$^+$ abundance based on H$_2$ column density derived from MAMBO 1.3 mm data for dust temperature, T$_{\rm KIN}$= 15 K.
[c]The N$_2$H$^+$ column density determined for excitation temperatures, $T_{\rm EX}$= 5 K.
[d]Fractional N$_2$H$^+$ abundance based on H$_2$ column density derived from MAMBO 1.3 mm data for dust temperature, T$_{\rm KIN}$= 15 K.